# Numerical simulation of electrically-driven flows using OpenFOAM

Francisco Pimenta [a] and Manuel A. Alves [b]

*CEFT, Departamento de Engenharia Química, Faculdade de Engenharia da Universidade do Porto, Rua Dr. Roberto Frias, 4200-465 Porto, Portugal*

In this work, we study the implementation of electrically-driven flow (EDF) models in the finite-volume framework of OpenFOAM®. The Poisson-Nernst-Planck model is used for the transport of charged species and it is coupled to the Navier-Stokes equations, governing the fluid flow. In addition, the Poisson-Boltzmann and Debye-Hückel models are also implemented. The discretization method and the boundary conditions are carefully handled in order to ensure conservativeness of ionic species in generic meshes and second-order accuracy in space and time. The applicability of the developed solver is illustrated in two relevant EDFs: induced-charge electroosmosis around a conducting cylinder and charge transport across an ion-selective membrane. The solver developed in the present work, freely available as open-source, can be a valuable and versatile tool in the investigation of generic electrically-driven flows.

___________________________________

[a] fpimenta@fe.up.pt
[b] mmalves@fe.up.pt

## I. INTRODUCTION

The interaction between fluids and electric fields at the micro- and nano-scales is at the origin of several phenomena with increasing interest. These interactions are frequently grouped in two classes, electrohydrodynamics (EHD) and electrokinetics (EK) [1,2], that we collectively call electrically-driven flows (EDFs) throughout this work. While EHD is characterized by the accumulation of net-charge at an interface, the distinguishing feature of EK is the formation of neutral electric double layers (EDLs) [1]. EK and EHD play an important role in fuel cells [3], electrodyalisis [4], separation techniques [5,6], fluid pumping [7,8] and mixing [8,9] in microfluidic systems, among other applications.

Understanding EK and EHD phenomena through theoretical models has been a challenge pursued over the last decades. Nevertheless, some questions remain unanswered and gaps still exist in the very basic theory [1,10]. As in other areas of physics, this is both due to the failure of the theory in capturing all the events involved – owing to simplifications or simply because they are unknown –, and to difficulties in extracting solutions from the available models. Regarding this last issue, closed form solutions to the system of equations that typically arises in EDFs are often difficult to be obtained, even for simple geometries. Matched asymptotes [11,12] and numerical methods [13-17] are among the approaches frequently adopted to solve the resulting system of equations, where the latter are particularly suitable to simulate EDFs in complex geometries.

Finite-differences [13, 18-25], finite-elements [13, 26-31], finite-volumes [16, 32, 33, 34-43] and lattice Boltzmann [15, 44-46] are numerical methods that have been successfully used to simulate EDFs. In what concerns available software packages, COMSOL Multiphysics® (https://www.comsol.com/), based on the finite-element method, has been a popular choice for this purpose [13, 26-29]. In the range of open-source packages, some works using Gerris [32, 33] and OpenFOAM® [34-39, 47-49] can be found in the literature, but none of these packages consistently offers a wide-range of ready-to-use models for EDFs, although both allow the user to build its own models. In this work, we use the OpenFOAM® package due to the following features: devoid of paid licenses, handling of generic polyhedral grids, parallel computation capability, possibility of creating/changing the source code, ease of integration between modules (rheological models, multiphase models, etc.) and wide acceptance among both the academic and industrial communities.

Concerning the simulation of EDFs with OpenFOAM®, Nandigana and Aluru [37] coupled the Poisson-Nernst-Planck equations to the Navier-Stokes equations to study the *I-V* curve in a micro-nanochannel. An area-averaged multi-ion transport model has then been derived to study similar problems [38, 39]. The Poisson-Nernst-Planck model, coupled to the Stokes equations, has also been used to study the flow inside a nanopore in steady conditions [47-49]. Zografos *et al* [36] optimized contraction-expansion microchannels to generate homogeneous extensional electroosmotic flows using the Debye-Hückel model, a simplification of the Poisson-Nernst-Planck model. In the range of two-phase EDFs, Lima and d'Ávila [34] implemented a leaky dielectric model to study the deformation of viscoelastic droplets under the action of electric fields. Roghair *et al* [35] used a similar model to analyze electrowetting of Newtonian fluids, where both the fluid and solid domains were numerically simulated in a coupled way. To the best of our knowledge, only the solver used in the work of Roghair *et al* [35] has been made available to the public, which, however, only implements the leaky dielectric model for two-phase EDFs of Newtonian fluids.

This work reports the numerical implementation of models to simulate single-phase EDFs of Newtonian fluids, using the OpenFOAM® toolbox. The models are implemented on the top of *rheoTool* (https://github.com/fppimenta/rheoTool), an open-source toolbox for the simulation of Generalized Newtonian/viscoelastic fluid flows using OpenFOAM®, which is now able to simulate both pressure- and electrically-driven flows (individually or mixed). We restrict this work to the analysis of electrokinetic problems, using the Poisson-Nernst-Planck, Poisson-Boltzmann and Debye-Hückel models, but other models (slip models, Ohmic model, among others) and additional cases can be found in *rheoTool*. The applicability of the solver is further demonstrated in the simulation of two particular EDFs: induced-charge electroosmosis (ICEO) around a conducting cylinder and charge transport across an ion-selective membrane. These cases are also used to assess the accuracy and stability of the numerical algorithm, and the results obtained can be used for benchmark purposes. Overall, this work is an attempt to increase the availability of general-purpose open-source solvers with built-in EDF models, for fluids of general rheology.

The remainder of this paper is organized as follows: Section II presents the governing equations and Section III describes the details of their numerical implementation in the finite-volume framework of OpenFOAM®. The performance of the numerical method in the application cases is discussed in Section IV. Finally, Section V presents the main conclusions and perspectives for future works.

## II. GOVERNING EQUATIONS

In the simulation of EDFs, two different, but coupled, components can be identified: the hydrodynamic component, represented by the continuity and momentum equations; the electric component, usually embodied by the Poisson-Nernst-Planck equations for the transport of charged species in dilute electrolytes. Both components are addressed in the next sections.

### A. Hydrodynamics: continuity and momentum equations

Consider the transient, incompressible, isothermal, single-phase, laminar flow of a Newtonian fluid under the action of an electric force. The continuity (Eq. 1) and momentum (Eq. 2) equations are

$$\nabla \cdot \mathbf{u} = 0 \tag{1}$$

$$\rho\left(\frac{\partial \mathbf{u}}{\partial t} + \mathbf{u} \cdot \nabla \mathbf{u}\right) = -\nabla p + \eta \nabla^2 \mathbf{u} + \mathbf{f}_E \tag{2}$$

where $\mathbf{u}$ is the velocity vector, $t$ is the time, $p$ is the pressure, $\mathbf{f}_E$ represents the electric force per unit volume, $\rho$ is the fluid density and $\eta$ is the viscosity.

The coupling between hydrodynamics and the electric force is ensured by the term $\mathbf{f}_E$. Ignoring magnetic effects, this term can be derived from the electrostatic Maxwell stress tensor, $\boldsymbol{\sigma} = \varepsilon\left(\mathbf{E}\mathbf{E} - \frac{\|\mathbf{E}\|^2}{2}\mathbf{I}\right)$, by taking its divergence

$$\mathbf{f}_E = \nabla \cdot \boldsymbol{\sigma} = \rho_E \mathbf{E} - \frac{\|\mathbf{E}\|^2}{2}\nabla \varepsilon \tag{3}$$

where $\varepsilon$ is the electric permittivity of the fluid, $\mathbf{E}$ is the electric field and $\rho_E$ is the charge density. The electric permittivity of the fluid is considered constant in all the problems addressed in this work, thus $\mathbf{f}_E = \rho_E \mathbf{E}$ in those cases. While the electric field is generically defined by the negative gradient of the electric potential, the computation of $\rho_E$ is model-dependent, as we show next.

### B. Poisson-Nernst-Planck equations

The transport of charged species (ions) in a low-ionic strength electrolyte, under the action of an electric field, can be described by the Nernst-Planck equation

$$\frac{\partial c_i}{\partial t} + \mathbf{u} \cdot \nabla c_i = \nabla \cdot (D_i \nabla c_i) + \nabla \cdot [(\mu_i \nabla \Psi)c_i] \tag{4}$$

where $c_i$ is the molar concentration, $D_i$ is the diffusion coefficient, $\mu_i$ is the electrical mobility and $\Psi$ is the electric potential, with $\mathbf{E} = -\nabla\Psi$ in electrostatics. The index $i$ represents each individual species in the electrolyte. The terms in the left hand-side of Eq. (4) represent the material derivative of $c_i$, the first term in the right hand-side is the flux of $c_i$ due to diffusion and the last term represents the transport of $c_i$ by the action of the electric field, also known as electromigration. This electromigration term is formally similar to a convective term, where an electromigration velocity (or flux, after discretization) can be identified, $\mathbf{u}_{M,i} = \mu_i \nabla\Psi$. All the electrolytes considered in this work follow the Nernst-Einstein relation, $\mu_i = D_i \dfrac{ez_i}{kT}$, where $z_i$ is the charge valence, $e$ is the elementary charge, $k$ is the Boltzmann constant and $T$ is the absolute temperature. However, for the sake of generality we will keep the notation for a generic $\mu_i$, since the Nernst-Einstein relation does not apply for several polyelectrolytes, such as for example DNA molecules [50].

The electric potential distribution in a given domain can be computed from Gauss' law (ignoring polarization)

$$\nabla \cdot (\varepsilon \nabla \Psi) = -\rho_E \qquad (5)$$

with the charge density defined as

$$\rho_E = F \sum_{i=1}^{m} z_i c_i \qquad (6)$$

where $F$ represents Faraday's constant and $m$ is the number of different ionic species. Eqs. (1)–(6) form the basic theory used to simulate most of EDFs. However, the electric component of this set of partial differential equations is frequently simplified or transformed in order to avoid numerical issues, as for example the numerical stiffness arising from the significantly different length and time scales which may co-exist, or simply to enable the derivation of closed-form solutions. Two of these simplified models are presented next. Hereafter, the Poisson-Nernst-Planck model, defined by Eqs. (4)–(6), is abbreviated as PNP.

### C. Poisson-Boltzmann model

The Poisson-Boltzmann model is a widely adopted simplification of the PNP model, and is based on the assumption that the ionic species follow a Boltzmann distribution [44, 51, 52],

$$c_i = c_{i,0} \exp\left[-\frac{\mu_i}{D_i}(\psi - \psi_0)\right] \tag{7}$$

where $c_{i,0}$ is a reference concentration of species $i$ for which $\psi = \psi_0$, and $\psi$ represents the intrinsic electric potential, as described later in Section 3.1. Eq. (7) can also be seen as the result of integrating the Nernst-Planck equation (Eq. 4) to a reference point (where $c_i = c_{i,0}$ and $\psi = \psi_0$), assuming a zero material derivative (steady-state and negligible transport of ions by convection). Variable $\Psi$ has been replaced by $\psi$ in order to ease the definition of $\psi_0$ in a generic situation where an externally applied electric potential co-exists with an intrinsic potential. This simplified model has a restricted applicability comparing to the generic PNP model, mainly due to the assumption of Boltzmann equilibrium to a fixed reference point. More details on these restrictions can be found elsewhere [52] but, roughly speaking, the Poisson-Boltzmann model is essentially valid for steady-state problems with negligible charge transport, non-overlapping EDLs and low intrinsic potentials.

We will assume $\psi_0 = 0$ henceforth, which is equivalent to consider electroneutrality in the reference point, usually the bulk solution. This assumption imposes no further restrictions other than the ones previously mentioned (different pairs $\{\psi_0, c_{i,0}\}$ would equally define Eq. 7), and is only taken to remove variable $\psi_0$ from the equation. When Eq. (7) is inserted in Eq. (6), the charge density of the so-called Poisson-Boltzmann model is obtained

$$\rho_E = F\sum_{i=1}^{m} z_i c_{i,0} \exp\left(-\frac{\mu_i}{D_i}\psi\right) \tag{8}$$

Eqs. (5) and (8) define the Poisson-Boltzmann model, hereafter abbreviated as PB, where the electric component is now decoupled from the hydrodynamics (but not the opposite). In addition, the only unknown regarding the electric component of the system is the electric potential, which is computed from Eqs. (5) and (8).

### D. Debye-Hückel model

The PB model can be further simplified to the so-called Debye-Hückel model in the limit of low electric potentials, $\left|\frac{\mu_i}{D_i}\psi\right| \ll 1$. Under this approximation, the charge density can be obtained by expanding the exponential term of Eq. (8) in a Taylor series up to the second term

$$\rho_E = F\sum_{i=1}^{m} z_i c_{i,0}\left(1 - \frac{\mu_i}{D_i}\psi\right) \tag{9}$$

Eqs. (5) and (9) define the Debye-Hückel model, hereafter abbreviated as DH. Again, the electric component is also decoupled from the hydrodynamics.

## III. NUMERICAL METHOD

The numerical discretization of the governing equations in the finite-volume framework follows the standard procedure available in OpenFOAM®, described elsewhere for related partial differential equations, e.g. in Moukalled *et al* [53] and Pimenta and Alves [54]. This methodology is applied in a consistent operator-basis for collocated grids. For the sake of conciseness, the discretization procedure for standard operators (time-derivatives, convective terms, etc.) is not presented here. In this section, we restrict our analysis to the particularities of EDFs.

### A. Splitting the electric potential into external and intrinsic electric potentials

When solving EDF problems with the simplified models (PB and DH), it is often convenient to decompose the electric potential in two variables, $\Psi = \psi + \phi_{Ext}$, where $\phi_{Ext}$ is the electric potential externally imposed and $\psi$ is the electric potential that exists intrinsically in the channel, commonly associated with the EDL formation [51]. Under this approach, Gauss' law is also decomposed into two equations

$$\nabla \cdot (\varepsilon \nabla \Psi) = -\rho_E \Rightarrow \begin{cases} \nabla \cdot (\varepsilon \nabla \phi_{Ext}) = 0 \\ \nabla \cdot (\varepsilon \nabla \psi) = -\rho_E \end{cases} \tag{10}$$

solved independently and sequentially for each variable. Following this strategy, it is also a common practice to only consider the contribution from $\phi_{Ext}$ in the computation of the electric field entering the electric force term of the momentum equation (Eq. 3), thus $\mathbf{f}_E = -\rho_E \phi_{Ext}$, as explained in Kyoungjin *et al* [51]. This is mainly to avoid the development of a large pressure gradient near the walls, due to the normal gradient of $\psi$ in these regions – this contribution can be canceled in the momentum equation by considering that it is balanced by an opposing pressure gradient, assuming that it does not influence the flow [51]. Such transformation can be particularly important (advantageous) in the case of viscoelastic fluids, where the development of large unbalanced pressure gradients may artificially trigger flow instabilities.

**B. Wall boundary conditions in EDFs**

Several materials used in EDFs at the micro- and nano-scales are impermeable and electrically insulating, as for example glass and polydimethylsiloxane (PDMS). In contact with an electrolyte, these materials develop electric double layers at the solid/liquid interface. The boundary conditions usually assigned to such surfaces, considering the PNP model, are a null velocity, a no-flux condition for ions and a specified electric potential or specified surface-charge (constant or variable over the surface).

The no-flux condition at the wall leads to (assuming no-penetration on the wall)

$$F_{i,f} = 0 \Rightarrow \left[ D_i \nabla c_i + c_i \mu_i \nabla \Psi \right]_f \cdot \mathbf{S}_f = 0 \qquad (11)$$

where $\mathbf{S}_f$ is the wall-normal vector, whose magnitude represents the area of face f lying on the wall. Eq. (11) is formally a Robin-type boundary condition for $c_i$. There are several options to convert Eq. (11) into a Neumann or Dirichlet boundary condition, for the ease of implementation, and three of these methods are as follows

**(I)** $\quad \nabla c_i \big|_f \cdot \mathbf{n} = -c_{i,f} \dfrac{\mu_i}{D_i} \nabla \Psi \big|_f \cdot \mathbf{n}$

**(II)** $\quad c_{i,f} = \dfrac{c_{i,P}}{1 + \dfrac{\mu_i}{D_i}(\Psi_f - \Psi_P)}$

**(III)** $\quad c_{i,f} = c_{i,P} \exp\left[ -\dfrac{\mu_i}{D_i}(\Psi_f - \Psi_P) \right] \quad$ and $\quad \nabla c_i \big|_f \cdot \mathbf{n} = -c_{i,f} \dfrac{\mu_i}{D_i} \nabla \Psi \big|_f \cdot \mathbf{n}$

with $c_{i,P}$ representing the value of $c_i$ evaluated at the center of the cell owning the boundary face where $c_{i,f}$ is to be computed (subscripts P and f represent the cell and the face, respectively), and $\mathbf{n} = \dfrac{\mathbf{S}_f}{|\mathbf{S}_f|}$ is the unitary vector, normal to face f. Method (I) results directly from Eq. (11) by simply isolating each term in a different side of the equation. Method (II) is obtained from method (I) after discretizing the gradients on the boundary and isolating all the $c_{i,f}$ terms. Method (III) derives an expression for $c_{i,f}$ based on the analytical integration of Eq. (11) between the cell and the face, which naturally results in an exponential variation of $c_i$ near the boundary. This exponential behavior is preserved when computing the gradient of $c_i$ by using the analytical expression for $c_{i,f}$, instead of taking $\nabla c_i \big|_f \cdot \mathbf{n} = \dfrac{c_{i,f} - c_{i,P}}{\delta}$, with $\delta$ being the distance between the cell center and face f. This last approach would generically linearize the exponential variation, leading to higher errors in coarse grids.

The three methods above are strictly similar and effective in imposing $F_{i,f} = 0$. Indeed, all satisfy equally Eq. (11) at the discrete level, which states that the diffusive and electromigration fluxes cancel each other at the boundary. This could be otherwise accomplished at the matrix construction stage by simply ignoring the contributions (at the no-flux boundary) from the diffusive and electromigration fluxes to the source vector, thus avoiding the need to compute $c_{i,f}$ or its gradient at the wall. In practice, the three methods above can only be distinguished based on the accuracy of the $c_{i,f}$ value retrieved when this variable is used for post-processing purposes, or when it is used in some high-order methods to increase the accuracy in non-orthogonal grids. In addition, the value of $c_{i,f}$ is also needed to compute limiters when using high-resolution schemes for the convective term of Eq. (4). Based on these criteria, method (III) was seen to be more accurate and less mesh-dependent in non-orthogonal grids. Thus, method (III) was used throughout this work. It should be noted that when using this method to evaluate $c_{i,f}$, the value of $c_{i,P}$ from the previous time-step/iteration is considered, resulting in an explicit boundary condition (an implicit implementation would be also a possibility).

Regarding the electric potential at the wall, as aforementioned the two common choices are either imposing the electric potential, or defining the surface charge density ($\sigma$), which can be related with the electric potential, $\nabla \Psi|_f \cdot \mathbf{n} = \frac{\sigma}{\varepsilon}$. For the PNP model, imposing the electric potential at the wall is a less natural boundary condition and it is also more complex to implement for walls of arbitrary shape, if a single electric potential variable is being used. A common workaround to this issue is to also decompose the electric potential in the PNP model.

When using the simplified PB and DH models, only the boundary conditions for the two electric potentials are required, since they are the only electric-related variables being solved. Considering the decomposition of the electric potential discussed in the previous section, the externally applied electric field is tangential to an insulating wall, while the intrinsic potential may be related with the zeta-potential ($\xi$) or with the surface charge density,

$$\begin{cases} \psi_f = \xi \text{ or } \nabla \psi|_f \cdot \mathbf{n} = \frac{\sigma}{\varepsilon} & \text{(intrinsic potential)} \\ \nabla \phi_{Ext}|_f \cdot \mathbf{n} = 0 & \text{(externally applied potential)} \end{cases} \quad (12)$$

In order to complete the discussion regarding the boundary conditions assigned at a wall, the pressure variable remains to be analyzed. A generic boundary condition can be derived for this variable from the pressure (continuity) equation. Indeed, recalling the SIMPLEC method described in our previous work [54], the velocity in a generic cell P can be written as

$$\mathbf{u}_P = \frac{\mathbf{H}}{a_P} + \mathbf{B} - \frac{(\nabla p)_P}{a_P - H_1} \tag{13}$$

which is the equation applied to correct the velocity after computing the pressure, being also used in the formulation of the pressure equation. In Eq. (13), $a_P$ represents the diagonal coefficients of the momentum equation and $H_1$ is the negative sum of all the off-diagonal coefficients. $\mathbf{H}$ is a vector containing the negative sum of the off-diagonal coefficients multiplied by the respective velocity and it also takes into account the source term contributions to the momentum equation, except the pressure gradient. The term $\mathbf{B} = \left( \frac{1}{a_P - H_1} - \frac{1}{a_P} \right)(\nabla p^*)_P$ is specific from the SIMPLEC algorithm, and more details can be found in Pimenta and Alves [54]. When Eq. (13) is interpolated to the faces of the domain, imposing a no-penetration condition at the wall ($\mathbf{u}_f \cdot \mathbf{n} = 0$) results in the following boundary condition for the pressure

$$(\nabla p)_f \cdot \mathbf{n} = (a_P - H_1)\left( \frac{\mathbf{H}_f}{a_P} + \mathbf{B}_f \right) \cdot \mathbf{n} \tag{14}$$

where both the $a_P$ and $H_1$ coefficients are from the cell owning the boundary face (only operators $\mathbf{H}$ and $\mathbf{B}$ are directly evaluated at the face). Eq. (14) has the advantage of being independent of the form taken by the momentum equation, since the contributions from all its terms are condensed in generic operators. Physically, Eq. (14) establishes a balance between the pressure and the remaining sources of momentum at the wall, in our case, viscous and electric stresses. A common approach in generic incompressible flows is to use $(\nabla p)_f \cdot \mathbf{n} = 0$, the so-called *zero-gradient* approach for pressure. This approximation does not violate continuity, but the pressure gradient at the boundary will not necessarily balance the remaining forces in the momentum equation, which can be of considerable magnitude in the case of EDFs, unless the normal electric stresses are artificially removed (balanced), as described in the previous section for the simplified models.

**C. Discretization of the electromigration term**

The electromigration term of the Nernst-Planck equation (Eq. 4) can be computed in different ways, which despite being equivalent analytically, might differ at the discrete level. One option is to split the term according to the properties of the divergence operator,

$$\nabla \cdot \left[(\mu_i \nabla \Psi) c_i\right] = \mu_i \left(c_i \nabla \cdot \nabla \Psi + \nabla c_i \cdot \nabla \Psi\right) \tag{15}$$

While the second term of Eq. (15) is likely to be discretized explicitly, the first term can be discretized semi-implicitly using Eqs. (5) and (6).

A second option is to handle the electromigration as a standard convective term, allowing for an implicit discretization. In this approach, the electromigration flux, $F_{M,i}|_f = (\mu_i \nabla \Psi)_f \cdot \mathbf{S}_f$, needs to be evaluated consistently on the cell faces in order to balance the remaining fluxes. This can be achieved by computing $\nabla \Psi|_f$, the negative of the electric field at cell faces, using the electric potential values straddling each face, instead of interpolating the gradients evaluated at the cell centers. The concentration variable, $c_i$, is then interpolated to face centers using an adequate scheme.

A third option is to simply consider the electromigration term as the Laplacian of the electric potential, with a variable coefficient $(\mu_i c_i)$ that can be linearly interpolated from cell centers to face centers. The implementation of this method requires an explicit evaluation of the electromigration term.

The last two methods were tested in the cases addressed in this work and no significant differences were observed between them in what respects accuracy and stability. Thus, the last method has been used due to its lower computational cost. Importantly, the no-flux boundary condition should be consistent with the discretization of the electromigration term, such that the fluxes exactly cancel out at the boundary.

**D. Linearization of the exponential terms in the Poisson-Boltzmann equation**

In order to improve the numerical stability of the PB model, the exponential source term included in Eq. (8) is linearized using a Taylor expansion up to the second term [55]. In its original form, the electric potential for this model is computed from

$$\nabla \cdot (\varepsilon \nabla \psi) = -F \sum_{i=1}^{m} z_i c_{i,0} \underbrace{\exp\left(\underbrace{-\frac{\mu_i}{D_i} \psi}_{b_i}\right)}_{a_i} \tag{16}$$

where the two variables $a_i$ and $b_i$ were introduced for the ease of notation in what follows. After linearization of the source term on the right hand-side of Eq. (16), the following equation is obtained

$$\nabla \cdot (\varepsilon \nabla \psi) + \psi F \sum_{i=1}^{m} (a_i b_i)^* = -F \sum_{i=1}^{m} (a_i)^* + \psi^* F \sum_{i=1}^{m} (a_i b_i)^* \tag{17}$$

where the terms with a star are evaluated explicitly. When the steady solution of Eq. (17) is reached, the second term in both sides of the equation exactly cancel each other and the original equation (Eq. 16) is recovered. In transient computations, the explicitness can be reduced by iterating Eq. (17) multiple times in order to update the terms with a star at each time-step.

**E. Discretization schemes**

The convective terms are discretized with the CUBISTA high-resolution scheme [56], following a deferred correction approach [54]. Both the Laplacian and gradient terms were discretized using central-differences. Overall, the algorithm is second-order accurate in space.

The time-derivatives were discretized using the three-time level scheme [54], rendering the algorithm also second-order accurate in time, as will be shown later.

All the terms of the momentum equation (Eq. 2), except the pressure gradient and the electric contribution, are discretized implicitly. In the Nernst-Planck equation (Eq. 4), only the electromigration term is accounted for explicitly, while in Poisson-type equations (either for pressure or for the electric potential) only the Laplacian operator is discretized implicitly, except in the Poisson equation for the DH model, where part of the $\rho_E$ term (Eq. 9) is also accounted for implicitly. Note that when mentioning implicit discretization, this excludes the corrective terms stemming, for example, from the deferred correction of convective terms (when using high-resolution schemes) or from the non-orthogonal correction of the Laplacian operator [53].

**F. Overview of the algorithm**

We use the same background solving sequence described in detail in Pimenta and Alves [54], which has been modified here to include the electric-related steps. In this segregated scheme, the coupling between the pressure and velocity fields is ensured by the SIMPLEC algorithm and an inner-iteration loop is used to reduce the explicitness of the method, and to increase its accuracy and stability. More details can be found in Pimenta and Alves [54]. Briefly, the sequence adopted in this work to solve EDFs consists of the following steps (noting that $\Psi$ should be replaced by $\phi_{Ext}$ and $\psi$ for the PB and DH models, which do not include $c_i$ as a computed variable):

1- Initialize the fields $\{ p, \mathbf{u}, \Psi, c_i \}_0$ and time ($t = 0$)
2- Enter the time loop ($t = \Delta t$)
   2.1- Enter the inner iterations loop (j = 0)

  2.1.1- Enter the electrokinetic coupling loop (k = 0)

    2.1.1.1. Compute $\Psi$ from Eq. (5) for the PNP model, or Eqs. (10) for PB and DH models

    2.1.1.2. Compute $c_i$ from the Nernst-Planck equation (Eq. 4) – skip this step for PB and DH models

    2.1.1.3. Increment the loop index (k = k + 1) and return to step 2.1.1.1 until the pre-defined number of coupling iterations is reached (only one iteration is needed for the PB and DH models)

  2.1.2- Solve the momentum equation (Eq. 2)

  2.1.3- Solve the pressure equation to enforce continuity (Eq. 1)

  2.1.4- Increment the inner iteration index (j = j + 1) and return to step 2.1.1 until the pre-defined number of inner iterations is reached

  2.1.5- Set { $p$, **u**, $\Psi$, $c_i$ }$_t$ = { $p_j$, **u**$_j$, $\Psi_j$, $c_{i,j}$ }

 2.2- Increment the time, $t = t + \Delta t$, and return to step 2.1 until the final time is reached

3- Stop the simulation and exit

From the above sequence, it is worth to mention the need of the so-called electrokinetic coupling loop (step 2.1.1). This loop is required to guarantee second-order accuracy in time for the PNP model, as will be shown later, and to enhance the coupling between the electric potential and the ionic concentration, reducing the non-linearity embodied by the electromigration term. This loop is not used with the PB and DH models, since those issues do not arise for these models.

The sparse matrices resulting from the discretization procedure are typically solved using a Pre-conditioned Bi-Conjugate Gradient (PBiCG) method coupled with a Diagonal Incomplete-LU (DILU) pre-conditioner, for non-symmetric matrices, and a Geometric-Algebraic Multi-Grid (GAMG) method coupled with a Diagonal Incomplete-Cholesky (DIC) pre-conditioner, for symmetric matrices. The absolute tolerance for the sparse-matrix solvers is typically set at $10^{-10}$.

## IV. RESULTS AND DISCUSSION

In this section, we start by assessing the convergence rate of the PNP model, both in space and time, before proceeding to the two selected application cases. We should note that all the cases addressed in this section are for symmetric, binary electrolytes, although the code developed is generic for any charge valence, diffusivity and number of species – the analysis of such generic cases is left as a suggestion for future work.

## A. Conservativeness, spatial and temporal accuracy of the PNP model: the 2D cavity

As shown in the previous section, several methods can be used to compute numerically the PNP equations. Even though some methods are analytically equivalent, they display different properties at the discrete level. The properties that we require for our discretized PNP equations are second-order accuracy in space and time, and conservation of the ionic species. Thus, it is convenient to first assess these points before advancing to more complex applications.

The problem selected for this purpose is the 2D cavity presented by Mirzadeh *et al* [18], which was also used by those authors to assess the conservation of ions by the PNP model in adaptive quadtree grids. It consists of a closed domain where an imposed sinusoidal variation of the electric potential along the walls generates a non-uniform distribution of ions (charge density). Due to the simple geometry (a 2D square) and conditions used, the 2D cavity is a good case to assess convergence rates and conservativeness for the PNP model. Here, we retrieve a new benchmark variable to this case, making it also affordable to probe the spatial accuracy of a numerical method.

### 1. Problem description

The cavity geometry is simply a square domain with side length $2H$ (Fig. 1). Although most of the tests were conducted using an orthogonal, structured-like mesh, some computations were also performed in a non-orthogonal mesh composed of triangles, as depicted in Fig. 1. In both types of meshes, the cells were compressed towards the domain boundaries, where the EDL develops (for the orthogonal mesh, the cell at each corner of the domain has a square shape). The minimum boundary edge size for the orthogonal meshes ranged between $H/800$ in mesh M1, and $H/4000$ in mesh M6, and the range was $H/1200$ in mesh M1N, and $H/4600$ in mesh M3N, for the non-orthogonal-meshes. The spatial resolution of all the meshes can be found in Fig. 2b.

The cavity is initially filled with a symmetric, binary electrolyte at uniform concentration $c_0$, for which $z_+ = -z_- = z$ and $D_+ = D_- = D$. The boundary conditions at the cavity walls are no-flux for both ionic species and

$$\Psi = \begin{cases} V_a \sin(\pi x/H) & y/H = 1 \\ -V_a \sin(\pi x/H) & y/H = -1 \\ V_a \sin(\pi y/H) & x/H = 1 \\ -V_a \sin(\pi y/H) & x/H = -1 \end{cases}$$

for the electric potential (this expression is slightly different from the one presented by Mirzadeh et al [18]). The hydrodynamics is switched off in this problem, such that the ions can only be transported by diffusion and electromigration.

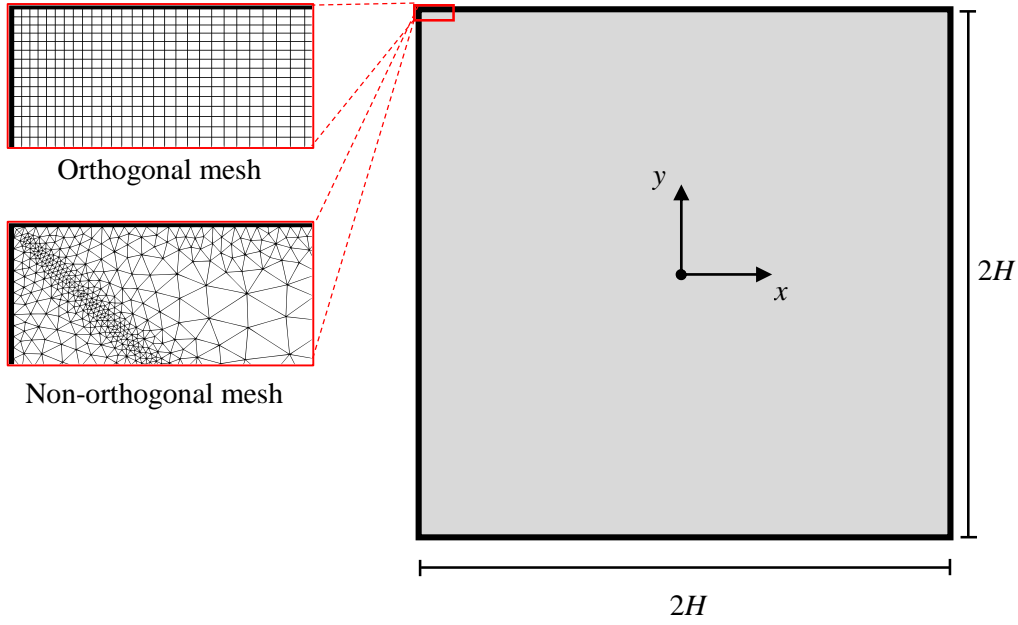

FIG. 1. Square domain of the two-dimensional cavity, with side length $2H$. The coordinate system is located at the center of the domain. A zoomed view near one corner of the orthogonal and non-orthogonal meshes is displayed next to the geometry.

The relevant dimensionless numbers in this problem are the dimensionless applied voltage, $\tilde{V}_a = \frac{V_a}{V_T}$, where $V_T = \frac{kT}{ez}$ is the thermal voltage, and the dimensionless Debye parameter, $\tilde{\kappa} = \frac{H}{\lambda_D} = \sqrt{\frac{2(Hz)^2 e c_0 F}{\varepsilon kT}}$, where $\lambda_D = \sqrt{\frac{\varepsilon kT}{2z^2 e c_0 F}}$ is the Debye length (an estimate of the EDL thickness). We follow Mirzadeh et al [18] setting $\tilde{V}_a = 5$ and $\tilde{\kappa} = 10$. The following dimensionless variables are used to display the results in this section:

$$\tilde{t} = \frac{tD}{H^2}, \ \tilde{y} = \frac{y}{H}, \ \tilde{x} = \frac{x}{H}, \ \tilde{\Psi} = \frac{\Psi}{V_T} \text{ and } \tilde{\rho}_E = \frac{(c_+ - c_-)}{c_0}.$$

*2. Results*

The contours of $\tilde{\Psi}$ and $|\tilde{\rho}_E|$ are plotted in Fig. 2a. The charge density is only non-negligible near the walls, where it displays a negative or positive value in the regions of positive or negative electric potential, respectively. The absolute charge density is symmetric in relation to the two diagonals of the square domain, as also in relation to the two Cartesian axes.

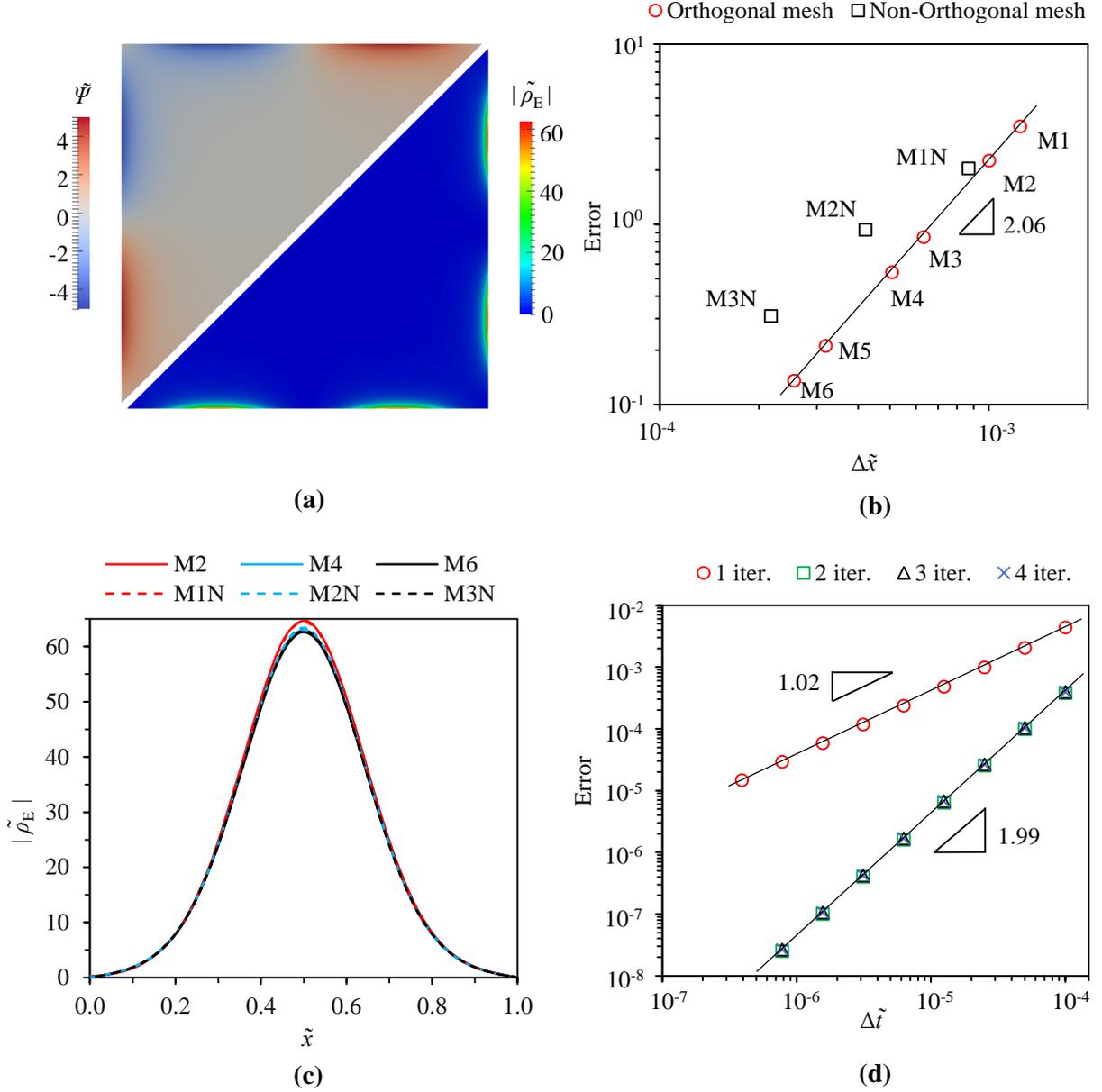

FIG. 2. (a) Contours of the electric potential (upper-diagonal) and absolute charge density (lower-diagonal) in mesh M6. (b) Spatial convergence rate with mesh refinement. The solid line represents a power-law fit to the numerical data obtained with the orthogonal mesh. In the $x$-axis we represent the minimum grid size in the direction normal to the cavity boundaries (edge size of the cell adjacent to the boundaries). (c) Spatial evolution of $|\tilde{\rho}_E|$ along the $x$-direction, at $\tilde{y} = 1$, for different meshes (see mesh numbering in panel b). The inset is a zoomed view of the profile near the peak region. (d) Temporal convergence rate for a different number of electrokinetic coupling iterations. The solid line represent a power-law fit to the numerical data. Mesh M4 was used to obtain these results (see mesh numbering in panel b).

We start assessing the spatial convergence rate using the peak values of $|\tilde{\rho}_E|$, at $(x, y)$ = $(\pm H/2, \pm H) \vee (\pm H, \pm H/2)$, which lie on the cavity boundaries (the local $|\tilde{\rho}_E|$ results from the no-flux boundary condition for the ions and from the imposed electric potential).

We estimate the variable in one of the eight locations by fitting a parabola to the three nearest face values lying on the boundary. The variable is computed in meshes of different resolution and Richardson's extrapolation to the limit is used to extrapolate the value for an infinitesimal cell spacing ($|\tilde{\rho}_E|_{\Delta x \to 0}$). Then, for each mesh resolution we compute the error as $\left| |\tilde{\rho}_E|_{\Delta x} - |\tilde{\rho}_E|_{\Delta x \to 0} \right|$. The results obtained are displayed in Fig. 2b, where we can confirm that the numerical method is second-order accurate in space. The non-orthogonal mesh presents a higher error than the orthogonal one, and its spatial accuracy seems to be approaching second-order for the two most refined meshes. The spatial variation of $|\tilde{\rho}_E|$ on the boundaries is plotted in Fig. 2c for different mesh resolutions, showing the convergence of the profiles (dependency on the mesh resolution is more clear near the peak region). Using the data from the orthogonal meshes, the Richardson's extrapolated value of $|\tilde{\rho}_E|$ at $(x, y) = (\pm H/2, \pm H) \vee (\pm H, \pm H/2)$ is 62.467.

In order to estimate the convergence rate in the temporal dimension, the value of $|\tilde{\rho}_E|$ was collected at $\tilde{t} = 0.001$, at the same position, i.e., $(x, y) = (\pm H/2, \pm H) \vee (\pm H, \pm H/2)$, using different time-steps and keeping the same mesh resolution. The extrapolation of that variable for an infinitesimal time-step was conducted as described above for the spatial convergence test and, for each time-step, the error was computed as $\left| |\tilde{\rho}_E|_{\Delta t} - |\tilde{\rho}_E|_{\Delta t \to 0} \right|$. The results are shown in Fig. 2d for a different number of iterations of the electrokinetic coupling loop (loop 2.1.1 in the algorithm described in section 3.6). For a single iteration, the method is only first-order accurate, notwithstanding the second-order accuracy of the discretization scheme employed for the time-derivatives. Second-order accuracy is only recovered for two or more coupling iterations. A similar behavior was described and explained in Karatay et al [13] for the PNP system of equations. We do not observed a significant improvement of accuracy by increasing from two to three or four iterations. Thus, two electrokinetic coupling iterations were typically used throughout this work.

As a final test, we computed the average change in the concentration of positive and negative ions for different time-steps, in both orthogonal and non-orthogonal meshes. Since no-flux boundary conditions are assigned to all domain boundaries, the average concentration of both ions must be conserved over time. For these specific tests, the absolute tolerance of the sparse-matrix solver of the Nernst-Planck equations was reduced to $10^{-14}$. The average concentration variation for each ionic species was quantified as $\Delta \tilde{c}_i = \left| \frac{1}{V_T} \sum_{k=1}^{NC} (\tilde{c}_{i,k} V_k) - 1 \right|$ where $V_T$ is the total volume of the domain, $V_k$ is the volume of cell k and NC is the total number of cells in the domain. The sum is carried out at $\tilde{t} = 2$, which is already in the steady-state regime of $\tilde{c}_i$. For a fully conservative method, $\Delta \tilde{c}_i = 0$ is expected. The range of time-steps tested spans two orders of magnitude, $\Delta \tilde{t} \in [0.0001, 0.01]$, and the values are significantly higher than the ones used to assess the temporal convergence rate, since higher losses are expected when using higher time-steps ($\Delta \tilde{t} = 0.01$ corresponds to one EDL charging time, $\lambda_D^2/D$). The results obtained are presented in Table I. The maximum loss observed was of $O(10^{-11})$, which confirms that the numerical method is conservative, both in orthogonal and non-orthogonal meshes. Furthermore, there is no clear relation between the mass loss and the time-step used. In one hand, a lower time-step leads to a more accurate evolution of the Nernst-Planck equations, reducing the numerical error. On the other hand, a lower time-step also requires more calculations until the steady-state is reached, which cumulatively deteriorates conservativeness due to round-off errors and also due to the finite tolerance of the iterative solvers.

TABLE I. Ionic species' mass variation using different time-steps and different meshes. Two electrokinetic coupling iterations were used for all the cases. See the definition of $\Delta \tilde{c}_i$ in the text and mesh numbering in Fig. 2b.

| $\Delta \tilde{t}$ | Mesh M4 | | Mesh M2N | |
|---|---|---|---|---|
| | $\Delta \tilde{c}_+$ | $\Delta \tilde{c}_-$ | $\Delta \tilde{c}_+$ | $\Delta \tilde{c}_-$ |
| 0.01 | 2.69 x $10^{-12}$ | 6.03 x $10^{-12}$ | 3.12 x $10^{-12}$ | 1.02 x $10^{-12}$ |
| 0.005 | 6.35 x $10^{-12}$ | 2.47 x $10^{-13}$ | 6.19 x $10^{-13}$ | 4.24 x $10^{-12}$ |
| 0.001 | 6.00 x $10^{-13}$ | 2.56 x $10^{-12}$ | 5.20 x $10^{-13}$ | 2.21 x $10^{-12}$ |
| 0.0005 | 1.12 x $10^{-11}$ | 5.25 x $10^{-13}$ | Not tested | Not tested |
| 0.0001 | 1.11 x $10^{-12}$ | 3.37 x $10^{-11}$ | Not tested | Not tested |

### B. Induced-charge electroosmosis (ICEO) around a conducting cylinder

The application case addressed in this section is the ICEO around a conducting cylinder, driven by a DC electric field. This EDF arises, for example, when a conducting (metallic) cylinder is placed in a DC (or AC) electric field. The potential induced on the cylinder surface drives 4 symmetric counter-rotating vortices, at low to moderate induced potentials [57]. Theoretically, the velocity scales with the square of the electric field magnitude (standard electroosmosis/electrophoresis only scale linearly). This feature, associated with the possibility of using AC electric fields to drive a unidirectional flow, attracted the interest of the scientific community on ICEO, and a number of practical applications in microfluidics already exist [26, 58, 59]. Thus, the numerical investigation of ICEO is a relevant subject, even more due to the well-known failure of the basic theory in capturing the velocity magnitude observed experimentally (e.g. [60, 61]).

Sugioka [42] derived analytical expressions for the ICEO around a conducting cylinder, considering approximate models for both low and high induced potentials. The author also used a hybrid finite-element/finite-volume numerical method to assess the accuracy of the theory developed. In what follows, we will compare our numerical results with the numerical and analytical results of Sugioka [42]. This case is suitable to test the accuracy and stability of the solver in non-orthogonal meshes.

### *1. Problem description*

The geometry consists of a 2D cylinder (radius $R$) centered in a square domain (Fig. FIG. 3), filled with a symmetric, binary electrolyte ($z_+ = -z_- = z$ and $D_+ = D_- = D$) at uniform concentration, $c_0$, and initially at rest. The electrodes, having symmetric electric potentials ($+V$ and $-V$), are placed on the north and south boundaries of the domain, spaced apart $L = 100R$ from each other, while the remaining boundaries are considered impermeable, insulating walls. The size of the bounding domain was selected long enough in order to guarantee that the solution is independent of this parameter. The domain was divided into 4 equal blocks in the azimuthal direction to build the mesh. For mesh M1, the mesh size on the cylinder surface is $ds = R/50$ and $dr = R/100$ (320 cells in the azimuthal direction and 80 cells in the radial direction). Mesh M2 was obtained from mesh M1 by doubling the number of cells of each block, in each direction. The cells were uniformly distributed in the azimuthal direction, and were compressed towards the cylinder surface in the radial direction.

The following set of boundary conditions was considered for the PNP model:

- Electrodes: $\Psi = \pm V$, $\nabla p \cdot \mathbf{n} = 0$, $\mathbf{u} = \mathbf{0}$, $c_i = c_0$;
- Cylinder surface: $\Psi = 0$, $\nabla p \cdot \mathbf{n}$ given by Eq. (14), $\mathbf{u} = \mathbf{0}$, $F_i = 0$;
- Insulating walls: $\nabla \Psi \cdot \mathbf{n} = 0$, $\nabla p \cdot \mathbf{n} = 0$, $\mathbf{u} = \mathbf{0}$, $F_i = 0$.

Note that by imposing a fixed ionic concentration at the electrodes (a reservoir-like condition), concentration polarization is not prone to happen at their surface, and we can consider our case similar to the unbounded problem of Sugioka [42], regarding that point.

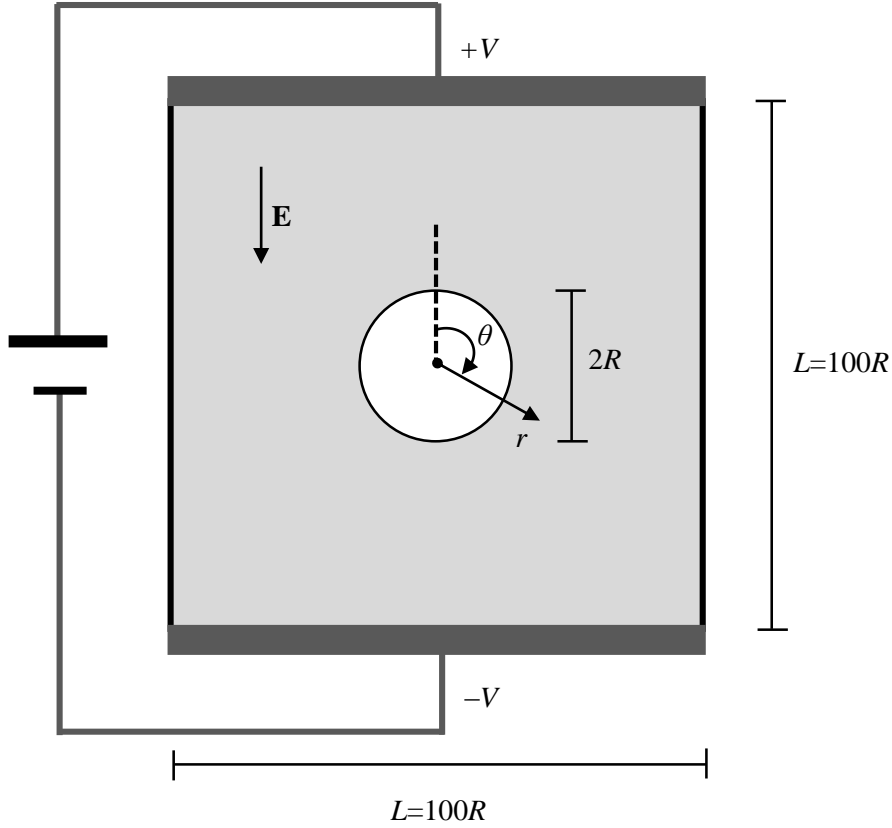

FIG. 3. Geometry used in the ICEO around a 2D conducting cylinder (drawing not to scale). The cylinder has radius $R$ and it is centered in a $100R$ square domain, composed of two electrodes having symmetric electric potentials (north and south boundaries) and two lateral impermeable, insulating walls (west and east boundaries). The cylinder has a fixed potential of 0 V.

The PB and DH models were also used in order to assess their accuracy against the PNP model, regarding the steady-state solution. For the PB and DH models, the following boundary conditions were used:

- Electrodes: $\psi = 0$, $\phi_{Ext} = \pm V$, $\nabla p \cdot \mathbf{n} = 0$, $\mathbf{u} = \mathbf{0}$;
- Cylinder surface: $\psi = -\phi_{Ext}$, $\nabla \phi_{Ext} \cdot \mathbf{n} = 0$, $\nabla p \cdot \mathbf{n} = 0$, $\mathbf{u} = \mathbf{0}$;
- Insulating walls: $\nabla \psi \cdot \mathbf{n} = 0$, $\nabla \phi_{Ext} \cdot \mathbf{n} = 0$, $\nabla p \cdot \mathbf{n} = 0$, $\mathbf{u} = \mathbf{0}$.

This set of boundary conditions ensures $\Psi = \psi + \phi_{Ext} = 0$ on the cylinder surface, i.e., the overall potential of the cylinder is kept constant and equal to its initial value.

A set of dimensionless numbers governing the ICEO are: dimensionless Debye parameter, $\tilde{\kappa} = \dfrac{R}{\lambda_D} = \sqrt{\dfrac{2(Rz)^2 ec_0 F}{\varepsilon kT}}$; Schmidt number, $Sc = \dfrac{\eta}{\rho D}$; electrohydrodynamic coupling constant, $\chi = \dfrac{\varepsilon V_T^2}{\eta D}$; dimensionless induced potential, $\tilde{V} = \dfrac{ER}{V_T}$, where $V_T = \dfrac{kT}{ez}$ is the thermal voltage and $E = \dfrac{2V}{L}$ is the applied electric field. To keep consistency with Sugioka [42], creeping flow conditions were imposed by removing the convective term of the momentum equation. In addition, $\chi = 0.47$ and $Sc = 10^3$, while both $\tilde{\kappa}$ and $\tilde{V}$ were varied in the range 5–10 and 0.01–4, respectively. The following dimensionless variables are used to present results: $\tilde{d} = \dfrac{r-R}{R}$, $\tilde{\mathbf{u}} = \dfrac{\mathbf{u}}{U}$ and $\tilde{\rho}_E = \dfrac{(c_+ - c_-)}{c_0}$, where $U = \dfrac{\varepsilon R E^2}{\eta}$ is a characteristic velocity scale for ICEO [11]. Note that here we use characteristic scales which are different from Sugioka [42] to avoid that some dimensionless numbers would depend on the size of the bounding domain.

*2. Results*

Fig. 4 presents radial profiles of the azimuthal velocity component, at $\theta = 45°$, for different combinations of parameters: $\tilde{V} = \{0.01; 4\}$ and $\tilde{\kappa} = \{5; 10\}$. An excellent agreement is observed in Fig. 4a,b between our numerical results in both meshes M1 and M2 and the analytical solution of Sugioka [42], for the low induced voltage ($\tilde{V} = 0.01$). For the higher induced voltage ($\tilde{V} = 4$), Fig. 4c,d show some discrepancies, although the qualitative behavior is still captured. The differences at $\tilde{V} = 4$ are possibly a consequence of the approximations at the basis of the high-voltage theory derived by Sugioka [42]. In general, our velocity profiles are in closer agreement with the analytical solution, than the numerical results of Sugioka [42], particularly far from the cylinder surface. This is most probably due to the longer domain used in the present work. The small domain used by Sugioka [42] ($L/R = 10$) was seen to influence the results in the current work, worsening the agreement with the analytical solution, which does not take into account wall effects.

The maximum azimuthal velocity component along the radial line $\theta = 45°$ is plotted in Fig. 5a as a function of the dimensionless Debye parameter, for low and high voltages. A good agreement is observed with the analytical and numerical results of Sugioka [42]. As the Debye parameter increases (the EDL thickness decreases), the dimensionless maximum azimuthal velocity increases. On the other hand, for fixed $\tilde{\kappa} = 10$, the dimensionless velocity decreases with increasing voltage, as illustrated in Fig. 5b for a high-voltage range, and previously in the velocity profiles of Fig. 4. In experimental

ICEO, it is well known that the *standard* theory [11], from which $U$ (velocity scale used to normalize **u**) arises, overestimates the measured velocity [60, 61], and Fig. 5b shows that the velocities computed numerically are also below the *standard* theory prediction, being consistent with the experimental observations. The same theory predicts a quadratic scaling of the velocity with the applied electric field, or equivalently $U \propto V^2$, but the numerical results suggest a scaling power lower than 2 (~1.6, see inset of Fig. 5b). Davidson *et al* [20] also found a weaker dependence between the velocity at the poles of the cylinder ($\theta = 0, 90$) and the electric field, comparing to the *standard* theory prediction. We should note, however, that the *standard* theory is essentially valid for $\tilde{\kappa} \gg 1$ and $\tilde{V} \ll 1$ [11].

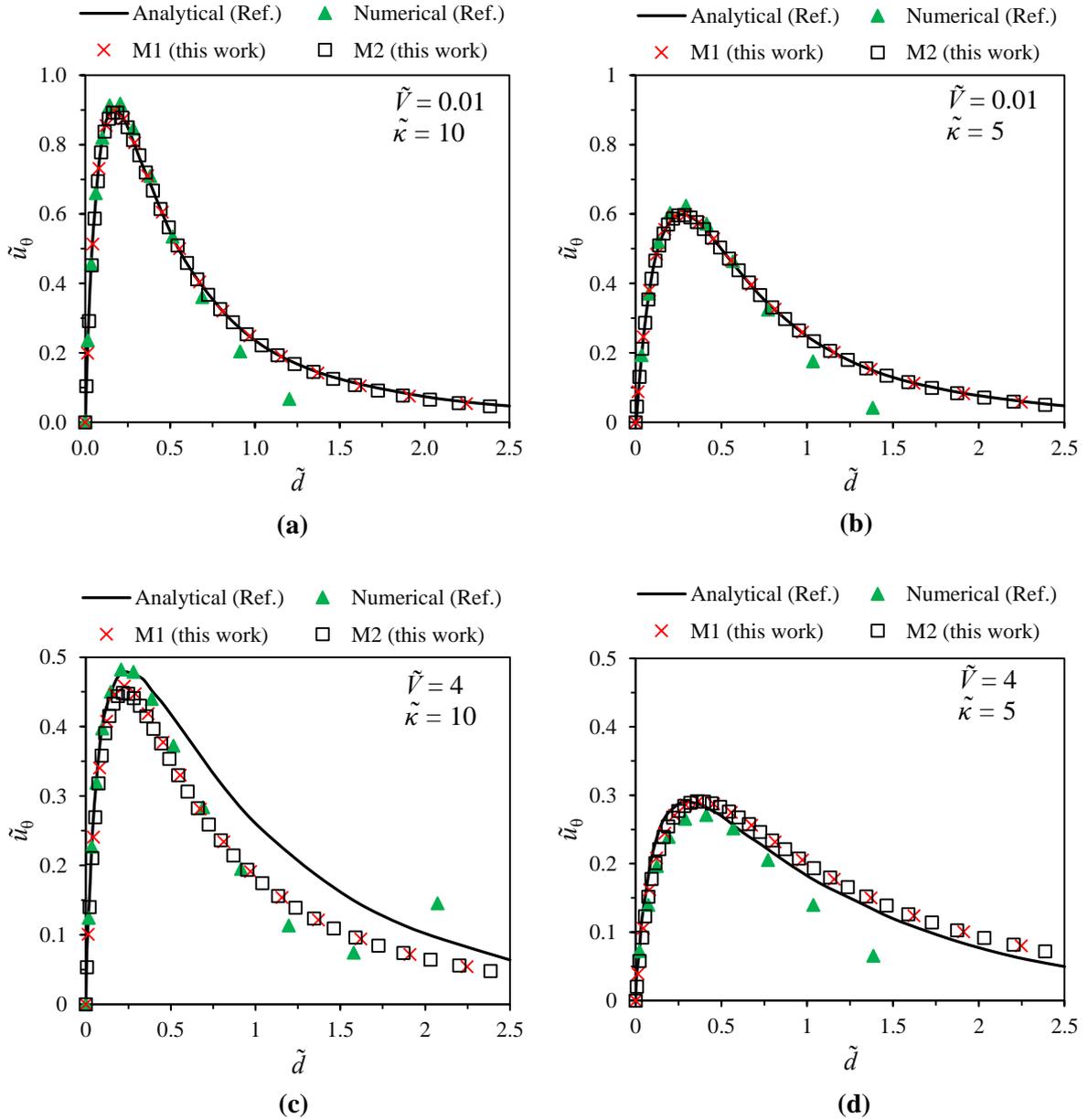

FIG. 4. Azimuthal velocity component profiles along the radial coordinate, at $\theta = 45°$: (a) $\tilde{V} = 0.01$ and $\tilde{\kappa} = 10$; (b) $\tilde{V} = 0.01$ and $\tilde{\kappa} = 5$; (c) $\tilde{V} = 4$ and $\tilde{\kappa} = 10$; (d) $\tilde{V} = 4$ and $\tilde{\kappa} = 5$. The analytical results for $\tilde{V} = 0.01$ and $\tilde{V} = 4$ are from the low-voltage and high-voltage theory for an unbounded domain, respectively, obtained from Sugioka [42]. The reference numerical data is also from that work.

The characteristic vortices of the ICEO flow are shown in Fig. 6, together with the charge density, for two different voltages $\tilde{V} = \{0.01; 4\}$, at $\tilde{\kappa} = 10$. While the velocity magnitude profiles are symmetric in relation to the lines $\theta = 0\text{-}180°$ and $\theta = 90\text{-}270°$, the charge density is only symmetric relative to line $\theta = 0\text{-}180°$, and it is anti-symmetric relative to line $\theta = 90\text{-}270°$. For the higher voltage ($\tilde{V} = 4$), a region of high velocity starts to form at $\theta = 0, 180°$ and the region of non-zero charge density is compressed towards the cylinder surface, which can only be seen in a zoomed view of Fig. 6.

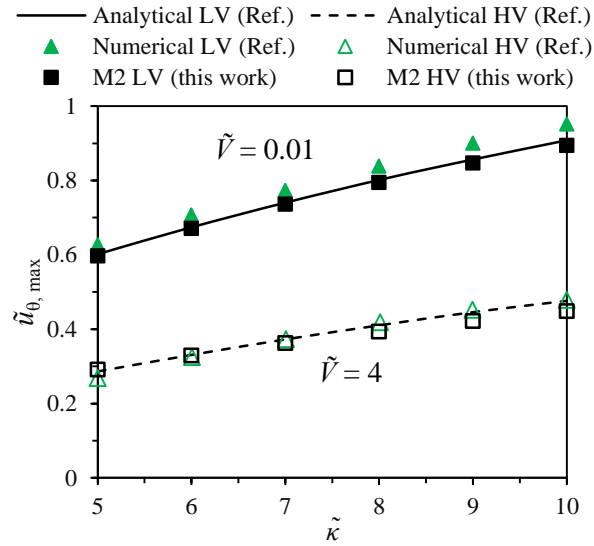

(a)

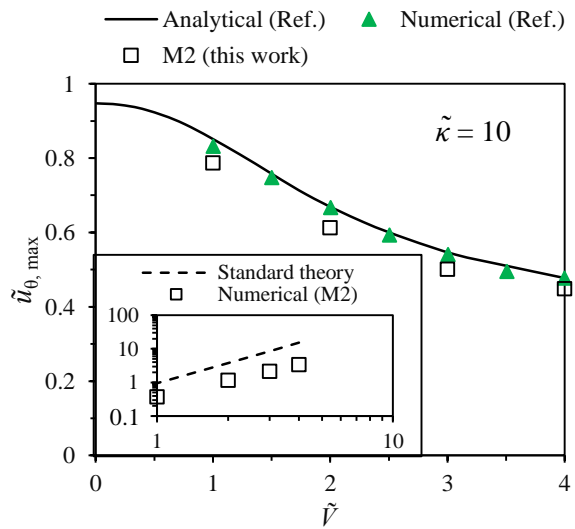

(b)

FIG. 5. Maximum azimuthal velocity along the radial line $\theta = 45°$ for (a) different $\tilde{\kappa}$ values, at $\tilde{V} = 0.01$ (LV – low-voltage) and $\tilde{V} = 4$ (HV – high-voltage), and (b) different applied potentials, at $\tilde{\kappa} = 10$. In panel (a),

the analytical results for $\tilde{V} = 0.01$ and $\tilde{V} = 4$ are from the low-voltage and high-voltage theory for an unbounded domain, respectively, as presented in Sugioka [42]. In panel (b), the curve with the analytical results is based on the high-voltage theory [42]. The reference numerical data is also from that work. The plot inset in panel (b) is a log-log representation of $|u_{\theta,\,max}|/(D/R)$ as a function of $\tilde{V}$. The dashed line represents the *standard* theory prediction: $u_{\theta=45,max} = 2\varepsilon RE^2/\eta$ [11], also normalized in the same way.

Fig. 7 presents a comparison between the velocity profiles obtained with the PNP, PB and DH models for high and low voltages. At low voltages, the two simplified models agree well with the Poisson-Nernst-Planck model, but a significant deviation is observed for the higher voltage, for which the assumptions at the basis of the two simplified models (essentially the assumption of Boltzmann equilibrium and $\tilde{V} \ll 1$) do not hold. Thus, using those two simplified models seems to be an acceptable approach at low voltages, since the accuracy is not compromised and the computations are usually faster (higher time-steps can be used due to the lower numerical stiffness and the CPU time per iteration is also smaller). This is important, for example, in shape optimization problems, where the steady-state solution has to be computed for each candidate geometry [36], and typically hundreds of geometries are tested until the optimal design is found, thus making the computational speed a key factor in these applications.

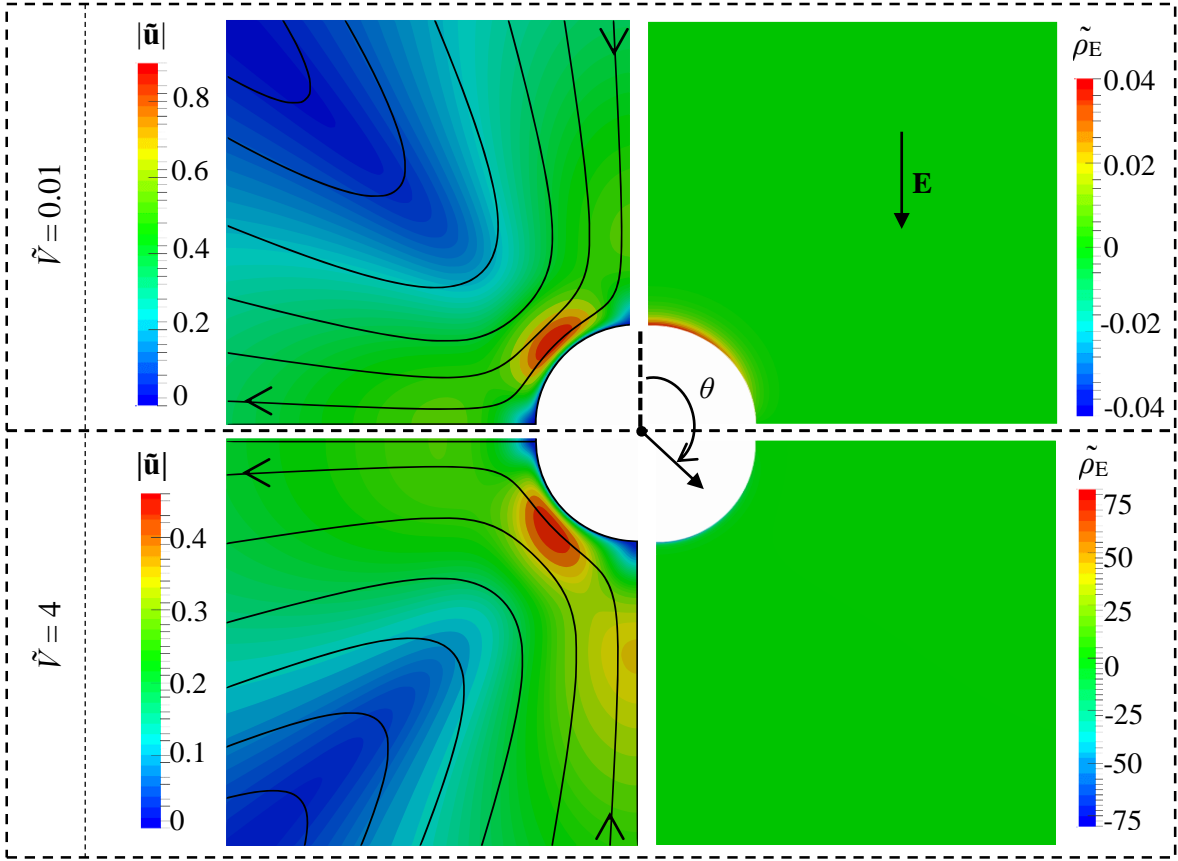

FIG. 6. Velocity contours with superimposed flow streamlines and charge density contours for $\tilde{V} = 0.01$ (top panels) and $\tilde{V} = 4$ (bottom panels). Each plot represents a different quadrant of the whole domain. All the results are for $\tilde{\kappa} = 10$ and were computed in mesh M2.

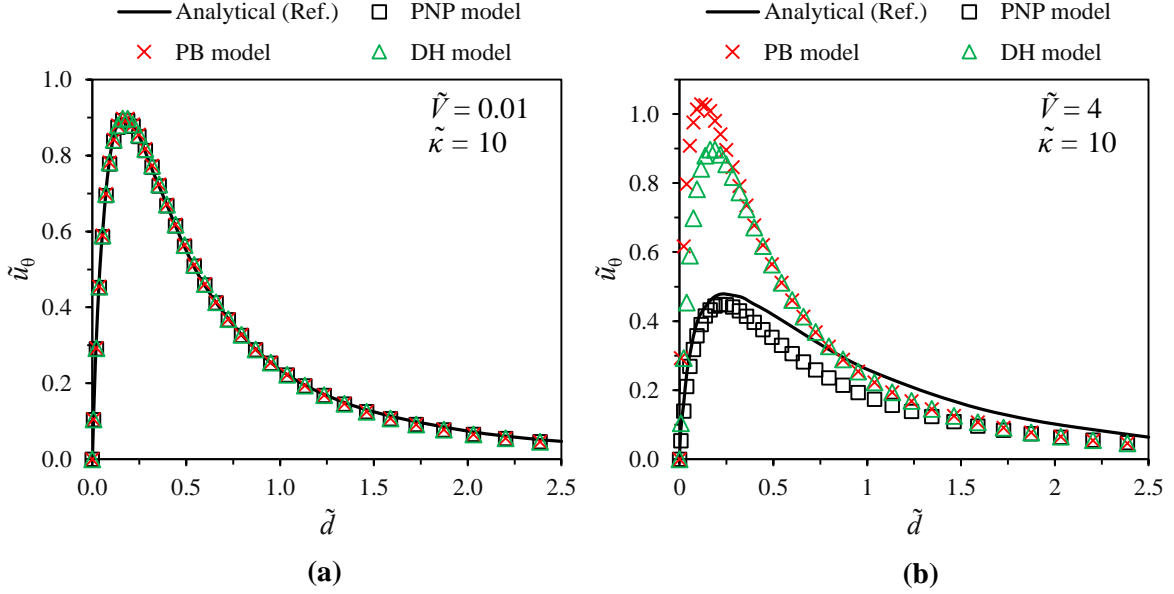

FIG. 7. Azimuthal velocity component profiles along the radial coordinate, at $\theta = 45°$, for (a) $\tilde{V} = 0.01$ and (b) $\tilde{V} = 4$, at $\tilde{\kappa} = 10$ (mesh M2), using different EDF models (PNP – Poisson-Nernst-Planck model; PB – Poisson-Boltzmann model; DH – Debye-Hückel model). The analytical results for $\tilde{V} = 0.01$ and $\tilde{V} = 4$ stem from the low-voltage and high-voltage theory in an unbounded domain, respectively, presented in Sugioka [42].

## C. Charge transport across an ion-selective membrane

When an electric potential difference is applied over an ion-selective membrane, for example in electrodialysis, the experimental *I-V* curve displays three regimes for increasing *V* [62, 63]: (i) Ohmic regime – *I* scales linearly with *V*; (ii) limiting regime – the rate of increase of *I* with *V* decays and *I* approaches an asymptotic value due to the diffusion-limited transport of ions; (iii) overlimiting regime – *I* increases again with *V*, although not necessarily in a linear way. The existence of an overlimiting regime has been commonly attributed to electroconvection – vortices form near the membrane and the resulting advection overcomes the diffusion-limited transport of ions. This has been observed experimentally [62-64] and predicted numerically [13, 21-25, 43, 65, 66], though direct numerical simulations (DNS) of this problem are still relatively recent and only a few are for 3D geometries [23, 25].

In this example, we use 2D DNS to obtain the *I-V* curve (or, equivalently, the *J-V* curve, with *J* being the current density) for an ion-selective membrane. This last application case is suitable to test the robustness and accuracy of the solver under the chaotic flow conditions developed at high voltages (electroconvective instabilities).

### 1. *Problem description*

The 2D reservoir considered in the present work (Fig. 8) has the same configuration and dimensions reported by Druzgalski *et al* [24], who simulated this problem using a second-order accurate (in time and space) finite-differences method. The distance between the electrode and the membrane is $H$, and $L = 6H$ is the length of the reservoir in the direction perpendicular to the applied electric field ($E = \Delta V/H$). An ion-selective membrane is located at $y = 0$, which allows the flux of cationic species, but retains anionic species. Periodic boundary conditions are imposed on the sides $x = \pm 0.5L$. We consider a symmetric, binary electrolyte, for which $z_+ = -z_- = z$ and $D_+ = D_- = D$.

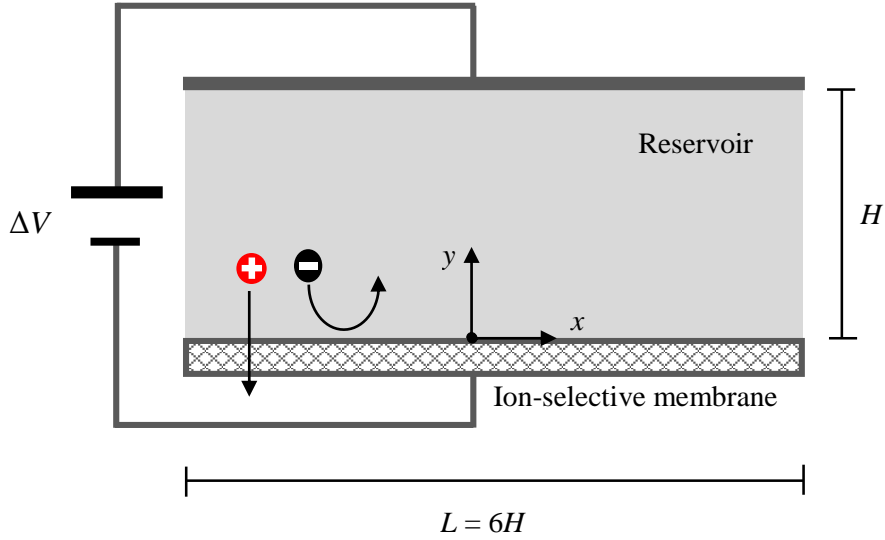

FIG. 8. Geometry representing a 2D reservoir with an ion-selective membrane at $y = 0$ (drawing not to scale). The reservoir width is $H$ and the length is $L = 6H$. Periodic boundary conditions are imposed on the sides, $x = \pm 0.5L$.

Two meshes were used in the numerical simulations in order to assess the convergence with mesh refinement. Mesh M1 has 480 uniformly spaced cells in the $x$-direction, and 90 cells non-uniformly distributed in the $y$-direction, with a minimum edge size of $H/1000$ near the ion-selective membrane. Mesh M2 was obtained from M1 by doubling the number of cells in each direction.

The following set of boundary conditions was used more details can be found in [24]:

- Reservoir ($y = H$): $\Psi = \Delta V$, $\nabla p \cdot \mathbf{n} = 0$, $\mathbf{u} = \mathbf{0}$, $c_i = c_0$, where $c_0$ is the initial concentration of ions in the bulk, in the absence of electric field;
- Periodic boundaries ($x = \pm 0.5L$): $\Omega|_{x=-0.5L} = \Omega|_{x=+0.5L}$, where $\Omega$ represents any of the computed variables;
- Ion-selective membrane ($y = 0$): $\Psi = 0$, $\nabla p \cdot \mathbf{n}$ given by Eq. (14), $\mathbf{u} = \mathbf{0}$, $c_+ = 2c_0$, $F_- = 0$.

In order to initialize the fields, the equivalent 1D problem is solved numerically in the absence of flow, keeping the same number of cells in the $y$-direction (the direction solved for), but only one single cell in the *x*-direction. The 1D solution is then mapped to the 2D domain and the anionic and cationic concentration fields are locally disturbed [24] by a 1 % random perturbation – the concentration in each cell is multiplied by a random scalar in the range [0.99;1.01].

The dimensionless numbers governing this EDF are similar to those of the ICEO case in the previous section: dimensionless Debye parameter, $\tilde{\kappa} = \dfrac{H}{\lambda_D} = \sqrt{\dfrac{2(Hz)^2 ec_0 F}{\varepsilon kT}}$; Schmidt number, $Sc = \dfrac{\eta}{\rho D}$; electrohydrodynamic coupling constant, $\chi = \dfrac{\varepsilon V_T^2}{\eta D}$; dimensionless potential, $\tilde{V} = \dfrac{\Delta V}{V_T}$, where $V_T = \dfrac{kT}{ez}$ is the thermal voltage. The following dimensionless variables are also used in this section: $\tilde{t} = \dfrac{tD}{H^2}$, $\tilde{y} = \dfrac{y}{H}$, $\tilde{x} = \dfrac{x}{H}$, $\tilde{\mathbf{u}} = \dfrac{\mathbf{u}}{U}$, $\tilde{c}_T = \dfrac{(c_+ + c_-)}{2c_0}$, $\tilde{\rho}_E = \dfrac{(c_+ - c_-)}{c_0}$ and $\tilde{J} = \dfrac{JH}{Dc_0 zF}$, where $J$ represents the current density and $U = \dfrac{D}{H}$ is a diffusive velocity scale. According to Druzgalski et al [24], we set $\chi = 0.5$, $Sc = 10^3$, $\tilde{\kappa} = 10^3$, $\tilde{V} = 10$–120 and creeping flow conditions are considered by removing the convective term of the momentum equation. One inner-iteration is used, along with two electrokinetic coupling iterations, and the time-step was fixed at $\Delta \tilde{t} = 10^{-6}$. The simulations were typically run until $\tilde{t} = 1$ (low voltages required more than this time to reach steady-state).

## 2. Results

The time evolution of the surface-averaged current density on the membrane ($J_{membrane}$) is plotted in Fig. 9a for $\tilde{V} = 10$–120 (unless otherwise stated, all the results presented in this section were computed in mesh M2). Since anions do not cross the membrane, the surface-averaged current density is computed from (using $\mu_i = D_i \dfrac{ez_i}{kT}$)

$$J_{membrane} = J_{+,membrane} = \dfrac{Fz_+}{L} \int_{membrane} -\left[ D_+ \nabla c_+ + \left( D_+ \dfrac{ez_+}{kT} \nabla \Psi \right) c_+ \right] \cdot \mathbf{n}\, dx \qquad (18)$$

For $\tilde{V} \leq 40$, the current density reaches a stationary value, either due to the negligible contribution of electroconvection ($\tilde{V} = 10$, not plotted in Fig. 9a), or due to the establishment of quasi-steady vortices ($20 \leq \tilde{V} \leq 40$). For $\tilde{V} > 40$, no steady-state is achieved and the current density profiles display a chaotic behavior over time for increasing $\tilde{V}$, as shown in Fig. 9a.

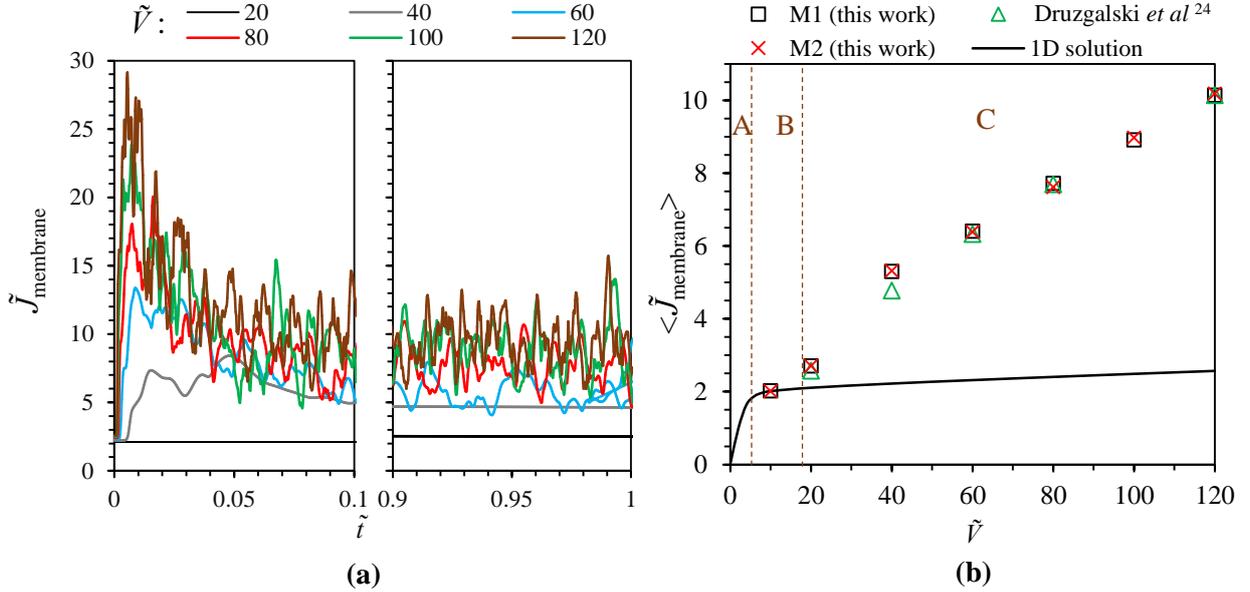

FIG. 9. (a) Time evolution of the surface-averaged current density on the membrane (mesh M2). (b) Space- and time-averaged current density on the membrane as a function of the dimensionless applied voltage. For $\tilde{V} > 40$, the average current density is computed by averaging the profiles plotted in panel (a) in the range $\tilde{t}$ = [0.2; 1], while for $\tilde{V} \leq 40$ the steady-state value is considered. The three regions of the J-V curve are identified (the delimitation between regions is only approximate): A – Ohmic regime; B – limiting regime; C – overlimiting regime.

The space- and time-averaged current density, $<J_{membrane}>$, is presented in Fig. 9b as a function of the dimensionless voltage. The transition from the limiting regime to the overlimiting regime occurs at $\tilde{V} \approx 20$. Additional simulations performed in mesh M1 showed that the transition occurs more precisely at $\tilde{V} = 19$ (results not shown), which agrees well with other works [21, 22, 24]. In general, a good agreement is observed with the results of Druzgalski et al [24], also plotted in Fig. 9b. The major deviations, either between our two meshes or between our results and the reference data, are observed at $\tilde{V} = 20$ and $\tilde{V} = 40$. The first voltage is in a sensitive region of transition between the limiting and overlimiting regimes, where a subcritical instability has been reported by several authors [21, 22, 43]. The reason for the discrepancy at $\tilde{V} = 40$ is probably due to the hysteresis of the J-V curve reported by Davidson et al [21] in the range $30 \leq \tilde{V} \leq 40$, although for a geometry with a higher aspect ratio (L/H). A vortex selection mechanism was pointed out as the main cause for this hysteresis [21].

Fig. 10 displays the x- and time-averaged kinetic energy in the y-direction, for different voltages (see the figure legend for details). The kinetic energy increases with the applied voltage, which makes the vortex-conveyed charge a very plausible explanation to the overlimiting regime. Furthermore, our profiles reproduce well those obtained by Druzgalski et al [24], notwithstanding the chaotic behavior observed at high voltages.

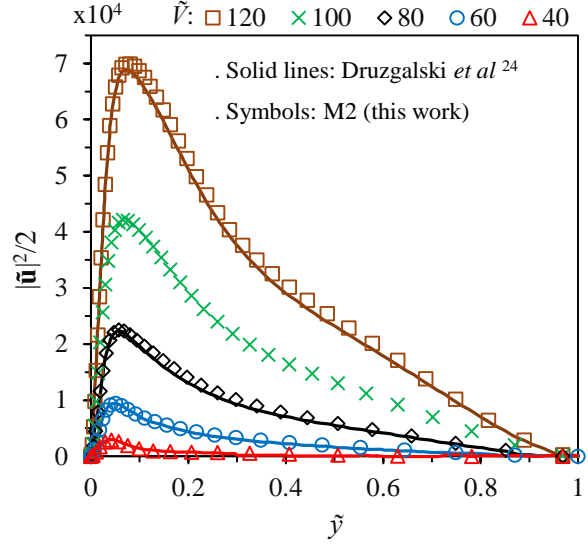

FIG. 10. Space-time-averaged profiles of the kinetic energy at different voltages. At each sampling time, the kinetic energy at each $y$-position was averaged along the $x$-direction. All the profiles collected over time were then averaged, for each $y$-position. The average in time was performed in the interval $\tilde{t} = [0.4; 1]$, over 20000 samples uniformly collected along that period. Symbols represent our numerical results, while lines correspond to the data in Druzgalski et al [24] (there is no available reference data for $\tilde{V} = 100$).

The chaotic patterns of the total ionic concentration and charge density at $\tilde{V} = 120$ are illustrated in Fig. 11 at different times. The mushroom-shaped zones of depleted fluid formed at early times quickly disrupt into random shapes, as also observed in Karatay et al [13]. For longer times, spikes of enriched fluid inject positive charge into the EDL near the membrane, while also removing negative charge from there. Interestingly, the adjacent strips of opposite charge reported by Druzgalski et al [24] are also present in Fig. 11. The chaotic nature at this voltage can be further assessed and measured by spectral analysis [13, 23, 24], a common approach in the study of turbulence. Therefore, we performed a spatial Fourier transform of the anionic concentration in the $x$-direction, at different distances from the membrane (for $\tilde{V} = 120$). The results are displayed in Fig. 12, where a power-law decay of the energy spectra is observed over almost one decade, at $\tilde{y} = 0.1$ and 0.4. The power-law exponent in this region is approximately -2, which is close to that obtained in Karatay et al [13] (estimated by visual inspection of their results). At $\tilde{y} = 0.8$, close to the reservoir boundary, the region of power-law decay is shorter, probably due to the reservoir boundary conditions and due to the finite range of action of the vortices (the instabilities are triggered and sustained near the membrane). The spectra for the cationic concentration are similar to those for anions (results not shown), except at $\tilde{y} = 0.025$, close to the membrane, where the spectrum for cations is much more energetic (by orders of magnitude) than that for anions – the near-membrane region is essentially depleted of anions, while cations are continuously injected on it by chaotic spikes. We have also confirmed the existence of a chaotic behavior over time, at different positions (results not shown).

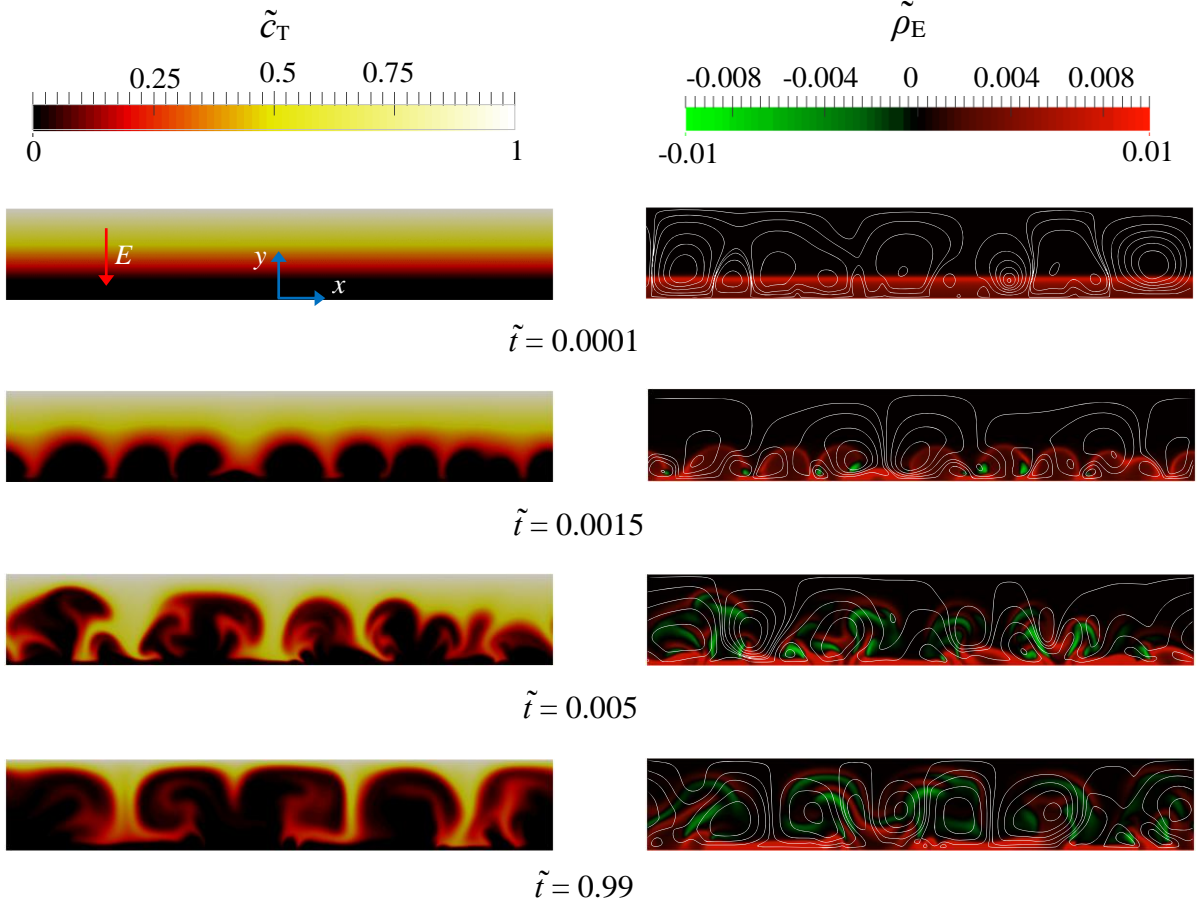

FIG. 11. Contours of the total concentration (left panel) and dimensionless charge density (right panel, with superimposed instantaneous streamlines) for $\tilde{V} = 120$, at different times. Note that the scale limits of the contours do not represent necessarily the real limits of the variable represented, in order to improve visibility. The whole domain is plotted.

A key aspect pointed out by Karatay *et al* [13] concerns the CPU time required to solve a given EDF problem. In their comparative study, the authors found speedup factors as high as one order of magnitude when using their in-house solver, in relation to the commercial COMSOL Multiphysics® package, for a case analogue to that addressed in this section [13]. This difference is not only due to the different numerical methods used (finite-differences vs. finite-elements), but it is also the result of a set of optimizations at the programming level, which can be done more easily in an in-house solver, but not in a general-purpose package because of generality reasons. In our case, using a mesh with 600 and 340 cells in the *x*- and *y*-direction (600×340×6 degrees of freedom), respectively, at $\tilde{V} = 120$, we obtained a CPU time of approximately 2 seconds per time-step in a laptop

i5-3210M processor (2.8 GHz, 3Mb cache), in a single-core run. In similar conditions, i.e., for the same degrees of freedom and same voltage, Karatay *et al* [13] reported a CPU time of approximately 1 second per time-step for their in-house solver and approximately 12 seconds per time-step using COMSOL Multiphysics®. Therefore, the OpenFOAM® solver used in this work also offers a good compromise between computational cost and generality (for example, in handling arbitrary geometries and grids).

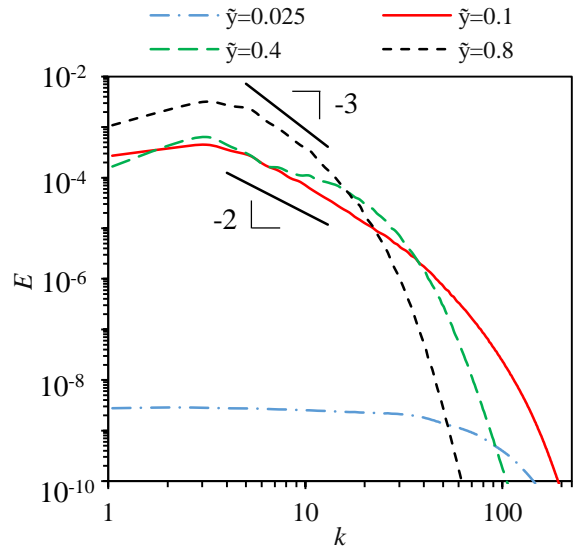

FIG. 12. Energy spectra of the anionic concentration variation along the *x*-direction, at different *y*-positions, for $\tilde{V} = 120$. In order to obtain representative data, the energy spectra were averaged over 20000 samples uniformly collected in the period $\tilde{t} = [0.4; 1]$. The wavenumber is $k = 2\pi\tilde{x}$.

## V. CONCLUSION

This work describes the numerical implementation of electrically-driven flow (EDF) models in *rheoTool*, an open-source toolbox to simulate flows of Generalized Newtonian and viscoelastic fluids using the finite-volume framework of OpenFOAM®.

Three EDF models were presented and discussed, including the Poisson-Nernst-Planck model and two simplifications, the Poisson-Boltzmann model and the Debye-Hückel model. After confirming the second-order accuracy in space and time, and the conservativeness of the Poisson-Nernst-Planck model, the developed solver was applied to two important EDFs: induced-charge electroosmosis around a conducting cylinder and charge transport across an ion-selective membrane. These application cases not only illustrated the applicability of the solver, as they also allowed to probe its accuracy and robustness in steady/transient and smooth/chaotic flow conditions. Furthermore, the numerical results obtained in this work increase the availability of benchmark data in non-trivial EDFs, for which exact analytical solutions are not available or are limited to a range of conditions.

A natural continuation of this work is the extension of EDFs to complex fluids, a feature already available in *rheoTool*, but not explored in the present work. This investigation is already under way and is left for future work. The study of EDFs in multiphase systems is also a relevant topic to be explored. In addition, a still relatively unexplored subject concerns EDFs with multiple species having different charge valences and diffusivities, which can be also simulated with *rheoTool*.

**ACKNOWLEDGEMENTS**

The research leading to these results has received funding from the European Research Council (ERC), under the European Commission "Ideas" specific programme of the 7$^{th}$ Framework Programme (Grant Agreement Nº 307499). The authors thank the relevant suggestions given by Professor Ali Mani.

[1] M. Z. Bazant, J. Fluid Mech. **782**, 1-4 (2015).
[2] A. Persat and J. G. Santiago, Curr. Opin. Colloid In. **24**, 52-63 (2016).
[3] G. Karimi and X. Li, J. Power Sources **140** (1), 1-11 (2005).
[4] S. S. Dukhin and N. A. Mishchuk, J. Membr. Sci. **79** (2), 199-210 (1993).
[5] K. D. Bartle and P. Myers, J. Chromatogr. **916** (1–2), 3-23 (2001).
[6] S. Ghosal, Electrophoresis **25** (2), 214-228 (2004).
[7] X. Wang, C. Cheng, S. Wang and S. Liu, Microfluid. Nanofluid. **6** (2), 145 (2009).
[8] A. Ramos, *Electrokinetics and electrohydrodynamics in microsystems*. (Springer Vienna, 2011).
[9] C.-C. Chang and R.-J. Yang, Microfluid. Nanofluid. **3** (5), 501-525 (2007).
[10] T. M. Squires, Lab. Chip. **9** (17), 2477-2483 (2009).
[11] T. M. Squires and M. Z. Bazant, J. Fluid Mech. **509**, 217-252 (2004).
[12] D. N. Petsev and G. P. Lopez, J. Colloid Interface Sci. **294** (2), 492-498 (2006).
[13] E. Karatay, C. L. Druzgalski and A. Mani, J. Colloid Interface Sci. **446**, 67-76 (2015).
[14] X. Luo, A. Beskok and G. E. Karniadakis, J. Comput. Phys. **229** (10), 3828-3847 (2010).
[15] H. Yoshida, T. Kinjo and H. Washizu, Commun. Nonlinear Sci. **19** (10), 3570-3590 (2014).
[16] H. S. Kwak and E. F. Hasselbrink, J. Colloid Interface Sci. **284** (2), 753-758 (2005).
[17] H. Sugioka, Colloids Surf. Physicochem. Eng. Aspects **376** (1–3), 102-110 (2011).
[18] M. Mirzadeh, M. Theillard and F. Gibou, J. Comput. Phys. **230** (5), 2125-2140 (2011).
[19] H. Liu and Z. Wang, J. Comput. Phys. **268**, 363-376 (2014).
[20] S. M. Davidson, M. B. Andersen and A. Mani, Phys. Rev. Lett. **112** (12), 128302 (2014).
[21] S. M. Davidson, M. Wessling and A. Mani, Sci. Rep. **6**, 22505 (2016).
[22] E. A. Demekhin, N. V. Nikitin and V. S. Shelistov, Phys. Fluids **25** (12), 122001 (2013).
[23] C. Druzgalski and A. Mani, Phys. Rev. Fluids **1** (7), 073601 (2016).
[24] C. L. Druzgalski, M. B. Andersen and A. Mani, Phys. Fluids **25** (11), 110804 (2013).
[25] E. A. Demekhin, N. V. Nikitin and V. S. Shelistov, Phys. Rev. E **90** (1), 013031 (2014).
[26] Z. Wu and D. Li, Microfluid. Nanofluid. **5** (1), 65-76 (2008).
[27] M. M. Gregersen, M. B. Andersen, G. Soni, C. Meinhart and H. Bruus, Phys. Rev. E **79** (6), 066316 (2009).
[28] C. P. Nielsen and H. Bruus, Phys. Rev. E **90** (4), 043020 (2014).
[29] H. Zhao and H. H. Bau, Langmuir **23** (7), 4053-4063 (2007).
[30] F. Bianchi, R. Ferrigno and H. H. Girault, Anal. Chem. **72** (9), 1987-1993 (2000).
[31] R. W. Lewis and R. W. Garner, Int. J. Numer. Meth. Eng. **5** (1), 41-55 (1972).
[32] C. Ferrera, J. M. López-Herrera, M. A. Herrada, J. M. Montanero and A. J. Acero, Phys. Fluids **25** (1), 012104 (2013).

[33] J. M. López-Herrera, S. Popinet and M. A. Herrada, J. Comput. Phys. **230** (5), 1939-1955 (2011).
[34] N. C. Lima and M. A. d'Ávila, J. Non-Newtonian Fluid Mech. **213**, 1-14 (2014).
[35] I. Roghair, M. Musterd, D. van den Ende, C. Kleijn, M. Kreutzer and F. Mugele, Microfluid. Nanofluid. **19** (2), 465-482 (2015).
[36] K. Zografos, F. Pimenta, M. A. Alves and M. S. N. Oliveira, Biomicrofluidics **10** (4), 043508 (2016).
[37] V. V. R. Nandigana and N. R. Aluru, J. Colloid Interface Sci. **384** (1), 162-171 (2012).
[38] V. V. R. Nandigana and N. R. Aluru, Electrochim. Acta **105**, 514-523 (2013).
[39] V. V. R. Nandigana and N. R. Aluru, Phys. Rev. E **94** (1), 012402 (2016).
[40] Y. Liu, L. Guo, X. Zhu, Q. Ran and R. Dutton, AIP Adv. **6** (8), 085022 (2016).
[41] A. M. Afonso, F. T. Pinho and M. A. Alves, J. Non-Newtonian Fluid Mech. **179–180**, 55-68 (2012).
[42] H. Sugioka, Phys. Rev. E **90** (1), 013007 (2014).
[43] V. S. Pham, Z. Li, K. M. Lim, J. K. White and J. Han, Phys. Rev. E **86** (4), 046310 (2012).
[44] M. Wang and Q. Kang, J. Comput. Phys. **229** (3), 728-744 (2010).
[45] D. Hlushkou, D. Kandhai and U. Tallarek, Int. J. Numer. Methods Fluids **46** (5), 507-532 (2004).
[46] G. H. Tang, Z. Li, J. K. Wang, Y. L. He and W. Q. Tao, J. Appl. Phys. **100** (9), 094908 (2006).
[47] M. Mao, J. D. Sherwood and S. Ghosal, J. Fluid Mech. **749**, 167–183 (2014).
[48] J. D. Sherwood, M. Mao and S. Ghosal, Phys. Fluids **26** (11), 112004 (2014).
[49] J. D. Sherwood, M. Mao and S. Ghosal, Langmuir **30** (31), 9261-9272 (2014).
[50] B. J. Kirby, *Micro- And Nanoscale Fluid Mechanics: Transport in Microfluidic Devices*. (Cambridge University Press, New York, 2010).
[51] K. Kyoungjin, K. Ho Sang and S. Tae-Ho, Fluid Dyn. Res. **43** (4), 041401 (2011).
[52] H. M. Park, J. S. Lee and T. W. Kim, J. Colloid Interface Sci. **315** (2), 731-739 (2007).
[53] F. Moukalled, L. Mangani and M. Darwish, *The finite volume method in computational fluid dynamics: an advanced introduction with OpenFOAM and Matlab*. (Springer Publishing Company, Incorporated, 2015).
[54] F. Pimenta and M. A. Alves, J. Non-Newtonian Fluid Mech. **239**, 85-104 (2017).
[55] S. Patankar, *Numerical heat transfer and fluid flow*. (Taylor & Francis, 1980).
[56] M. A. Alves, P. J. Oliveira and F. T. Pinho, Int. J. Numer. Methods Fluids **41** (1), 47-75 (2003).
[57] M. Z. Bazant and T. M. Squires, Phys. Rev. Lett. **92** (6), 066101 (2004).
[58] F. Zhang and D. Li, Electrophoresis **35** (20), 2922-2929 (2014).
[59] J. S. Paustian, A. J. Pascall, N. M. Wilson and T. M. Squires, Lab. Chip. **14** (17), 3300-3312 (2014).
[60] C. Canpolat, S. Qian and A. Beskok, Microfluid. Nanofluid. **14** (1), 153-162 (2013).
[61] J. A. Levitan, S. Devasenathipathy, V. Studer, Y. Ben, T. Thorsen, T. M. Squires and M. Z. Bazant, Colloids Surf. Physicochem. Eng. Aspects **267** (1–3), 122-132 (2005).
[62] S. Nam, I. Cho, J. Heo, G. Lim, M. Z. Bazant, D. J. Moon, G. Y. Sung and S. J. Kim, Phys. Rev. Lett. **114** (11), 114501 (2015).
[63] S. M. Rubinstein, G. Manukyan, A. Staicu, I. Rubinstein, B. Zaltzman, R. G. H. Lammertink, F. Mugele and M. Wessling, Phys. Rev. Lett. **101** (23), 236101 (2008).
[64] J. C. de Valença, R. M. Wagterveld, R. G. H. Lammertink and P. A. Tsai, Phys. Rev. E **92** (3), 031003 (2015).
[65] E. A. Demekhin, V. S. Shelistov and S. V. Polyanskikh, Phys. Rev. E **84** (3), 036318 (2011).
[66] V. V. Nikonenko, V. I. Vasil'eva, E. M. Akberova, A. M. Uzdenova, M. K. Urtenov, A. V. Kovalenko, N. P. Pismenskaya, S. A. Mareev and G. Pourcelly, Adv. Colloid Interface Sci. **235**, 233-246 (2016).